\documentclass[prd,aps,twocolumn,preprintnumbers,amsmath,amssymb,nofootinbib,superscriptaddress,notitlepage]{revtex4-1}

\usepackage{epsfig}
\usepackage[utf8]{inputenc}
\usepackage{color}
\usepackage{graphicx}
\usepackage{epstopdf}
\usepackage{float}
\usepackage[colorlinks=true,
linkcolor=blue,
breaklinks=true,
urlcolor=blue,
citecolor=blue]{hyperref}
\usepackage{multirow}

\newcommand{\nn}{\nonumber}
\newcommand{\be}{\begin{equation}}
\newcommand{\ee}{\end{equation}}
\newcommand{\bea}{\begin{eqnarray}}
\newcommand{\eea}{\end{eqnarray}}
\newcommand{\ba}{\begin{array}}
\newcommand{\ea}{\end{array}}
\newcommand{\bi}{\begin{itemize}}
\newcommand{\ei}{\end{itemize}}

\newcommand{\mcb}{{\mathcal B}}

\newcommand{\difd}{\mathrm d}

\newcommand{\lf}{\left}
\newcommand{\rg}{\right}

\newcommand{\ucas}{\affiliation{School of Nuclear Sciences and Technology, University of Chinese Academy of Sciences, Beijing 100049, China}}

\newcommand{\ynu}{\affiliation{Department of Physics, Yunnan University, Kunming 650091, China}}

\newcommand{\csr}{\affiliation{Research Center for Hadron and CSR Physics, Lanzhou University and Institute of Modern Physics of CAS, Lanzhou 730000, China}}

\newcommand{\keylab}{\affiliation{State Key Laboratory of Heavy Ion Science and Technology, Institute of Modern Physics, Chinese Academy of Sciences, Lanzhou 730000, China}}

\begin{document}

\title{Disentangling isospin violation in charmonia decaying into $\Xi \overline \Xi$ pairs}

\author{Huan-Ran Wen} %\email{wenhuanran@impcas.ac.cn}
\keylab
\ucas

\author{Chun-Qiu Zhao}\email{corresponding author: zhaochunqiu@stu.ynu.edu.cn}
\keylab
\ynu

\author{Xu Cao}\email{corresponding author: caoxu@impcas.ac.cn}
\keylab
\ucas
\csr

\author{Jian-Ping Dai}\email{corresponding author: daijianping@ynu.edu.cn}
\ynu

\date{\today}

\begin{abstract} 
  \rule{0ex}{3ex}
  Based on a model-independent isospin decomposition, we perform a data-driven analysis of $J/\psi$ and $\psi(2S)$ decays into $\Xi \bar{\Xi}$ pairs by leveraging the full set of observables from high-statistics BESIII data. We find that the isospin-violating amplitudes are predominantly driven by the electric couplings, with magnitudes reaching approximately 10\%, contrasting with the smaller magnetic contributions of about 3\%. A partial-wave analysis further reveals that $S$-wave isospin violation is comparable to the $D$-wave contribution. 
\end{abstract}

\maketitle

\section{Introduction} \label{sec:intro}

The decays of charmonium states into baryon and their antiparticles are of particular interest from both experimental and theoretical perspectives. A systematic description of those exclusive processes can be achieved using a perturbative approach within the QCD factorization framework~\cite{Bolz:1997as,Kivel:2021uzl,Kivel:2022fzk}. However, final-state rescattering interactions are essential to accurately account for the experimental data~\cite{Chen:2006yn,Geng:2026vkz}.
A large relative phase, nearly orthogonal, is found between the strong and electromagnetic amplitudes for both $J/\psi$ and $\psi(2S)$ decays~\cite{Zhu:2015bha}.
These processes can also be used to test the gluon spin~\cite{Brodsky:1981kj}, as well as to investigate constituent quark mass effects~\cite{Carimalo:1985mw}.
Regarding the testing of \textit{SU}(3) flavor symmetry~\cite{LopezCastro:1994xw,Wei:2009zzh,Mo:2021asa,Mo:2023wrf}, the breaking effect is found to be less significant for $\psi(2S)$ than for $J/\psi$~\cite{BaldiniFerroli:2019abd,Ferroli:2020mra}.
Even more intriguing is the mechanism of isospin violation stemming from up and down quark mass differences and electromagnetic effects, which remains a key issue in understanding these decays.
The isospin violations due to quark masses is expected to be much smaller than \textit{SU}(3) violations predominantly due to the strange-quark mass.
Early on, it was pointed out that isospin violations from electromagnetic corrections could be significant and depend sensitively on the baryon magnetic form factors~\cite{Claudson:1981fj}.

Noticeable isospin violations are not unique to baryonic final states but are also manifest in other exclusive channels. 
A notable example is $J/\psi \to K^*(892) \bar K$, where a very pronounced isospin breaking from the electromagnetic interaction was reported~\cite{BESIII:2026aig}.
The BESIII Collaboration has measured precisely the branching fraction for the isospin-violating decay $\psi(3770) \to J/\psi \pi^0$~\cite{BESIII:2012uuq,BESIII:2026pyf}, thereby enabling a reliable extraction of the light-quark mass ratio.
Further searches were conducted for the isospin-violating transitions $\chi_{c0,2} \rightarrow \pi^{0} \eta_{c}$ \cite{BESIII:2015osb} and doubly OZI-suppressed decay $J/\psi \to \phi \pi^0$~\cite{BESIII:2015asx}.
The $G$-parity violating decay $J/\psi \to \pi^+ \pi^-$ attracts interest as the one-photon exchange process is expected to dominate over other amplitudes that involve gluons and are hence suppressed~\cite{BaldiniFerroli:2016mbs}. 

The study of hyperons is especially advantageous because they act as self-analyzing polarimeters through their weak decays \cite{Lee:1957qs}. 
By measuring the polarization and correlations of the hyperon-antihyperon system, one can unambiguously determine the relative phase between the electric and magnetic form factors \cite{Faldt:2017kgy,Salone:2022lpt,Cao:2024tvz,BESIII:2018cnd,BESIII:2021ypr,BESIII:2024nif}. 
The isospin-violating decay $\psi \to \Sigma^0 \bar{\Lambda} + c.c.$, which is mediated by a purely electromagnetic mechanism and results in a branching fraction about two orders of magnitude lower than those of isospin-allowed channels \cite{BESIII:2012xdg,BESIII:2023cvk, BESIII:2021mus,ParticleData:2026uvj}, provides a reference scale for the magnitude of isospin-violating amplitudes as well as a unique opportunity to study two-photon exchange effects in the timelike region \cite{Cao:2026TPE}.
This work presents a model-independent separation of the electric and magnetic isospin-violating amplitudes in Sec.~\ref{sec:isodecomp}, and equivalently their $S$- and $D$-wave components in Sec.~\ref{sec:sdwave}, based on the full experimental information from the sequential decays of $J/\psi$ and $\psi(2S)$ into $\Xi\bar{\Xi}$ pairs.

\section{Isospin decomposition of amplitudes} \label{sec:isodecomp}

\begin{table*}[tbh]
\vspace{-2mm}
\centering
\caption{The isospin violation relevant ratio, $R_I^M$, and $R_I^E$ extracted from the branching ratios and the $\alpha_Y$. Earlier measurements are listed for reference but are now regarded as superseded.}
\label{tab:charmonium}
\begin{tabular}{lcccccc}
\hline\noalign{\smallskip}
Decay process & Branching ratio  & $\alpha_Y$ & $\Delta\Phi /$rad &  $|R_I^M|$ &  $|R_I^E|$ \\
\noalign{\smallskip}\hline\noalign{\smallskip}
\hline
$J/\psi \to \Xi^0 \overline \Xi^0$ & $(1.17 \pm 0.04) \times 10^{-3}$ \cite{ParticleDataGroup:2022pth} & 0.514$\pm$0.006$\pm$.0015 \cite{BESIII:2023drj} \footnote{0.66$\pm$0.03$\pm$0.05 \cite{BESIII:2016nix} is superseded.}& 1.168$\pm$0.019$\pm$0.018 \cite{BESIII:2023drj} &  \multirow{2}{*}{0.08$\pm$0.04} & \multirow{2}{*}{0.18$\pm$0.05} \\ % $3.42 \%$
$J/\psi \to \Xi^- \overline \Xi^+$ & $(0.97 \pm 0.08) \times 10^{-3}$ \cite{ParticleDataGroup:2022pth} & 
0.586$\pm$0.012$\pm$0.010 \cite{BESIII:2021ypr} \footnote{0.58$\pm$0.04$\pm$0.08 \cite{BESIII:2016ssr} is superseded.} & 1.213$\pm$0.046$\pm$0.016 \cite{BESIII:2021ypr} & & \\ %$8.25 \%$
\noalign{\smallskip}\hline\noalign{\smallskip}
$\psi(2S) \to \Xi^0 \overline \Xi^0$ & $ (2.59 \pm 0.02) \times 10^{-4}$ \cite{BESIII:2025dke} \footnote{$(2.3 \pm 0.4) \times 10^{-4}$ \cite{ParticleDataGroup:2022pth} is superseded} & 0.768$\pm$0.029$\pm$0.025 \cite{BESIII:2025dke} \footnote{0.65$\pm$0.09$\pm$0.14 \cite{BESIII:2016nix} and 0.665$\pm$0.086$\pm$0.081 \cite{BESIII:2023lkg} are superseded.} & 0.257$\pm$0.061$\pm$0.009 \cite{BESIII:2025dke} \footnote{-0.05$\pm$0.150$\pm$0.020 \cite{BESIII:2023lkg} is superseded.}& \multirow{2}{*}{0.040$\pm$0.032} & \multirow{2}{*}{0.19$\pm$0.16} \\
$\psi(2S) \to \Xi^- \overline \Xi^+$ & $(2.87 \pm 0.11) \times 10^{-4}$ \cite{ParticleDataGroup:2022pth} & 0.693$\pm$0.048$\pm$0.049 \cite{BESIII:2022lsz} \footnote{0.91$\pm$0.13$\pm$0.14 \cite{BESIII:2016ssr} is superseded.} & 0.667$\pm$0.111$\pm$0.058 \cite{BESIII:2022lsz} & & \\ %$8.25 \%$
\noalign{\smallskip}\hline\noalign{\smallskip}
\hline
\end{tabular} 
\end{table*}

It is feasible to perform a model-independent isospin decomposition of the timelike electromagnetic form factors for baryon isospin multiplets by transforming from the physical basis to the isospin basis \cite{Dai:2023vsw,Cao:2021asd}.
Similarly, the electromagnetic couplings in charmonium decays into baryon-antibaryon pairs can be decomposed into isospin components, where the isovector component accounts for isospin violation. 
The isospin decomposition for an isospin-doublet baryon pairs involves four independent complex amplitudes (2 isospin states $\times$ 2 electromagnetic couplings), yielding seven relevant real parameters after fixing a global phase. 
These parameters are solely determined by eight independent observables available at each energy point: the decay rates and angular distributions, along with the relative phases between electric and magnetic couplings derived from polarization observables for both members of the isospin doublet.
While applicable to both the nucleon and $\Xi$ isospin doublets, the latter is particularly advantageous as all relevant observables are accessible with high precision through self-analyzing weak decays. This facilitates a purely data-driven investigation of isospin violated amplitudes of charmonium, as detailed below.

We first recall the basic formula for the decay of a vector charmonium into an octet baryon-antibaryon pair $Y \bar{Y}$, where the angular distribution is given as
\be\label{eq:angdis}
\frac{\difd \Gamma_{Y} }{\difd \cos\theta} = \frac{M_{\psi} \beta}{32 \pi} \lf( |G_M|^2 + \frac{|G_E|^2}{\tau} \rg) (1+\alpha_{Y}\cos^2\theta) \,,
\ee
where $\theta$ denotes the angle between the baryon momentum and the positron beam direction in the center-of-mass (c.m.) frame. 
%(\textbf{coincidence with BESIII}).
$\beta = \sqrt{1 - \tau^{-1}}$ is the velocity of the baryon with $\tau = M_\psi^2/4M_Y^2$. The angular distribution parameter $\alpha_{Y}$ is given by $\alpha_{Y} = (\tau - R_Y^2)/(\tau + R_Y^2)$ with $R = |G_E/G_M|$. 
The partial decay branching ratio is
\bea \label{eq:width}
\mcb_{\psi \to Y \bar{Y}} &=& \frac{M_{\psi}\beta}{12 \pi \Gamma_{\psi}}\lf( |G_M|^2 + \frac{|G_E|^2}{2\tau} \rg) \nn \\ &\equiv& \frac{M_{\psi}\beta}{12 \pi \Gamma_{\psi}} \frac{2\tau +1}{2\tau} |G_{\textrm{eff}}|^2 
\eea
with $\Gamma_{\psi}$ being the total width of charmonium.
Thus, simultaneous measurements of the decay rates and angular distributions unambiguously separate the moduli of the electric and magnetic couplings.
In the case of unpolarized electron and positron beams, the transverse polarization of the hyperon and the spin correlation of hyperon-anti-hyperon are:
\bea \label{eq:Py} 
P_y &=& \frac{\sin2\theta }{\sqrt{\tau }D} \mathfrak{Im}(G_{M}G_{E}^*) \nn \\ &=& \frac{ \sqrt{1-\alpha_\psi^2} {\sin\!\theta \cos\!\theta}}{1+\alpha_{\psi}\cos^2\!\theta} \sin\Delta\Phi \,, \\
C_{xz} &=& \frac{\sin2\theta }{\sqrt{\tau }D}  \mathfrak{Re}(G_{M}G_{E}^*) \nn \\ &=& \frac{ \sqrt{1-\alpha_\psi^2} {\sin\!\theta \cos\!\theta}}{1+\alpha_{\psi}\cos^2\!\theta} \cos\Delta\Phi \,,
\label{eq:Cxz}
\eea
with $D=(1+\cos^2\theta)|G_{M}|^2+\frac{1}{\tau } \sin^2\theta |G_{E}|^2 $.
Those polarized observables can be reconstructed via the sequential decay chains of the hyperons \cite{Tomasi-Gustafsson:2005svz,Faldt:2017kgy,Cao:2024tvz}.
Consequently, the relative phases $\Delta \Phi$ between the electromagnetic couplings can be unambiguously determined, providing the necessary input for the isospin analysis.

We then perform an isospin decomposition of the electromagnetic couplings into their isoscalar$(I=0)$ and isovector $(I=1)$ components \cite{Cao:2021asd}:
\be \label{eq:isoamp}
G_{E,M}^{+,-} = \frac{G_{E,M}^1 \pm G_{E,M}^0}{2}
\ee
where $+$ and $-$ correspond to $\psi \rightarrow \Xi^{0}\bar{\Xi}^{0}$ and $\Xi^{-}\bar{\Xi}^{+}$ channels, respectively.
The complex amplitudes are parameterized, without loss of generality, as $G^{I}_{E,M} = |G^{I}_{E,M}| \, e^{i \varphi_{E,M}^I}$.
The strength and relative phase of the isospin-violating effect is characterized by
\be \label{eq:dI} 
\delta_{E,M} = \lf | \frac{G_{E,M}^1}{G_{E,M}^0} \rg | \,, \quad \phi_{E,M} = \textrm{arg} \lf [ \frac{G_{E,M}^1}{G_{E,M}^0} \rg] \,,
\ee
for electric and magnetic couplings, respectively.
Here the relative phases between the isovector and isoscalar components are defined as $\phi_{E,M} = \varphi_{E,M}^1 - \varphi_{E,M}^0$.
Since charmonium states are isospin singlets, any non-vanishing value of $\delta_{E,M}$ quantifies the degree of isospin violation in the decay process, with $\delta_{E,M} \ll 1$ being generally expected. Eq. \eqref{eq:isoamp} yields a practical ratio to bound isospin violation~\cite{Dai:2023vsw}:
\be \label{eq:dataRIExi}
R_I^{E,M} = \frac{|G_{E,M}^{+}|^2 - |G_{E,M}^{-}|^2}{|G_{E,M}^{+}|^2 + |G_{E,M}^{-}|^2} = \frac{2\delta_{E,M}\cos{\phi_{E,M}}}{1+\delta_{E,M}^2}
\ee
if the electric and magnetic couplings $|G_{E,M}^{\pm}|$ for both $\psi \rightarrow \Xi^{0}\bar{\Xi}^{0}$ and $\Xi^{-}\bar{\Xi}^{+}$ channels are individually measured.

Using Eqs.~\eqref{eq:width} and \eqref{eq:isoamp}, the isoscalar and isovector components can be conveniently expressed in terms of the unpolarized observables, $|G_{\textrm{eff}}|$ and $R_{\pm}$~\cite{Dai:2023vsw}:
\begin{widetext}
\bea 
|G^{0}_{M}|^{2}+|G^{1}_{M}|^{2} &=&\frac{2(2\tau_{+}+1)|G^{+}_{\textrm{eff}}|^{2}}{2\tau_{+} + R_{+}^{2}} + \frac{2(2\tau_{-}+1)|G^{-}_{\textrm{eff}}|^{2}}{2\tau_{-} + R_{-}^2}\label{eq:GMsum} \,, \\ 
|G^{0}_{E}|^{2}+|G^{1}_{E}|^{2} &=& \frac{2(2\tau_{+}+ 1) R_{+}^{2} |G^{+}_{\textrm{eff}}|^{2}}{2\tau_{+} + R_{+}^{2}} + \frac{2(2\tau_{-} + 1)R_{-}^{2}|G^{-}_{\textrm{eff}}|^{2}}{2\tau_{-} + R_{-}^{2}} \label{eq:GEsum} \,,\\ 
\mathfrak{Re} [ G^{1}_{M}  G^{0 \dag}_{M} ] &=& \frac{(2\tau_{+} + 1)|G^{+}_{\textrm{eff}}|^{2}}{2\tau_{+} + R_{+}^{2}} - \frac{(2\tau_{-} + 1)|G^{-}_{\textrm{eff}}|^{2}}{2\tau_{-} + R_{-}^2} \label{eq:GMsub}\,,\\ 
\mathfrak{Re} [ G^{1}_{E}  G^{0 \dag}_{E}] &=& \frac{(2\tau_{+} + 1)R_{+}^{2} |G^{+}_{\textrm{eff}}|^{2}}{2\tau_{+} + R_{+}^{2}} - \frac{(2\tau_{-} + 1)R_{-}^{2}|G^{-}_{\textrm{eff}}|^{2}}{2\tau_{-} + R_{-}^{2}} \label{eq:GEsub}\,,
\eea
\end{widetext}
Here, $\tau_{\pm}$ takes into account the small isospin-breaking effects arising from the $\Xi^{0,-}$ mass difference.
Achieving a complete separation of the electric and magnetic couplings in the isospin basis requires the knowledge of the relative phases $\Delta \Phi$ of both channels. 
This is achieved by substituting Eq.~\eqref{eq:isoamp} into the polarization observables in Eqs.~\eqref{eq:Py} and \eqref{eq:Cxz}, yielding:
\begin{widetext}
\bea
\mathfrak{Re} \lf[ {G^{1}_M}  G^{1*}_E + G^{0}_M G^{0*}_E \rg ] &=& 2 |G^+_M||G_E^+|\cos\Delta\Phi^+ + 2 |G_E^-||G_M^-|\cos\Delta\Phi^- \label{eq:EM1}
\\
\mathfrak{Re} \lf[ {G^{1}_M}  G^{0*}_E + G^{0}_M G^{1*}_E \rg ] &=& 2 |G^+_M||G_E^+|\cos\Delta\Phi^+ - 2 |G_E^-||G_M^-|\cos\Delta\Phi^- \label{eq:EM2}
\\
\mathfrak{Im} \lf[ {G^{1}_M}  G^{1*}_E + G^{0}_M G^{0*}_E \rg ] &=& 2 |G_M^+||G_E^+|\sin\Delta\Phi^+ + 2 |G_M^-||G_E^-|\sin\Delta\Phi^- \label{eq:EM3}
\\
\mathfrak{Im} \lf[ {G^{1}_M}  G^{0*}_E + G^{0}_M G^{1*}_E \rg ] &=& 2 |G_M^+||G_E^+|\sin\Delta\Phi^+ - 2 |G_M^-||G_E^-|\sin\Delta\Phi^- \label{eq:EM4}
\eea  
\end{widetext}
An alternative expression in terms of the polarization observables $P_y$ and $C_{xz}$ can be found in Appendix A of Ref.~\cite{Dai:2023vsw} \footnote{We note that there are typographical errors in Eqs.~(A.6) and (A.7) of Ref.~\cite{Dai:2023vsw}, where $\mathfrak{Re}$ and $\mathfrak{Im}$ should be interchanged.}.
This complete set of relations in Eqs.~\eqref{eq:GMsum}--\eqref{eq:EM4} enables the separation of the electric and magnetic couplings in the isospin basis ($I=0, 1$), transformed from the physical basis ($\Xi^0, \Xi^-$).

\begin{figure}
    \centering
    \includegraphics[width=0.75\linewidth]{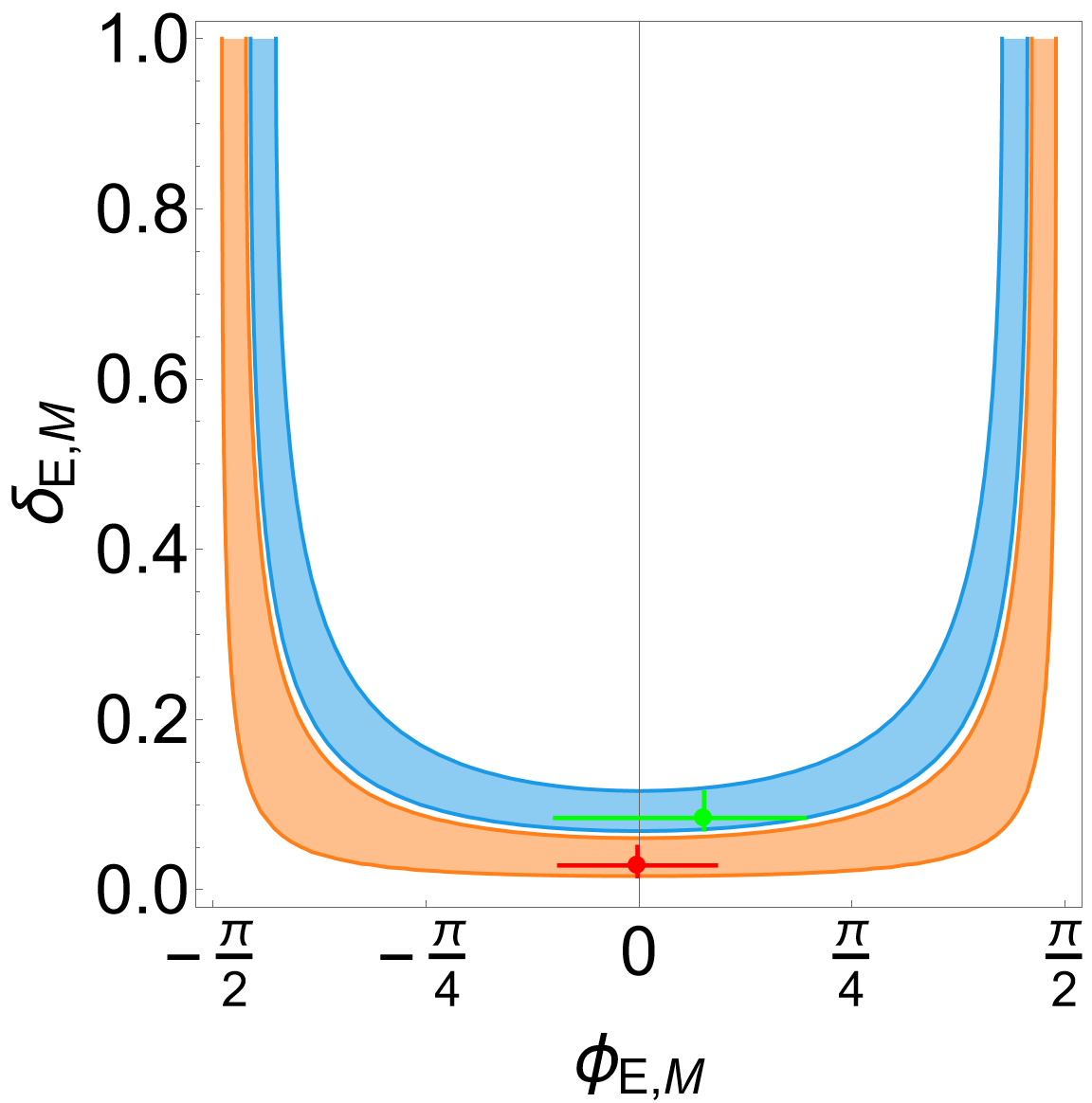}
    \caption{Allowed regions for $\delta_{E,M}$ and $\phi_{E,M}$ in $J/\psi\rightarrow\Xi\bar{\Xi}$ decays. The blue and orange bands represent constraints derived from the $R_I^E$ and $R_I^M$ values listed in Table~\ref{tab:charmonium}, respectively. The points with error bars denote the numerical solutions obtained from the $\chi^2$ minimization.}
    \label{fig:RIEMofJPsi}
\end{figure}

\begin{figure}
    \centering
    \includegraphics[width=0.75\linewidth]{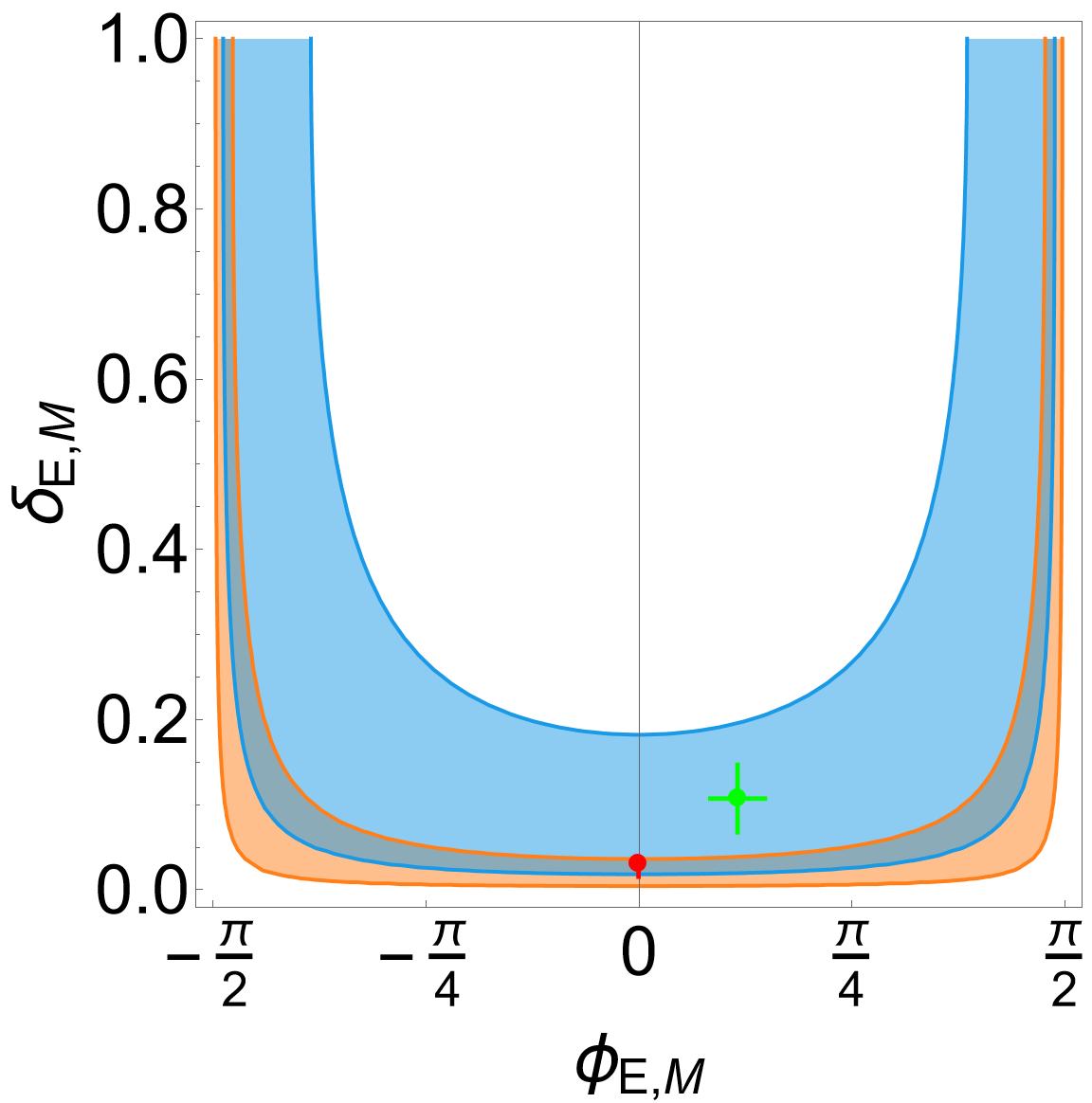}
    \caption{Same as Fig. \ref{fig:RIEMofJPsi} but for $\psi(2S) \rightarrow \Xi\bar{\Xi}$ decay.}
    \label{fig:RIEMofPsi2S}
\end{figure}

Two subtleties should be noted when applying them to practical use.
First, the global phase is arbitrary, and we fix $\varphi_{M}^{0} = 0$ by convention. Consequently, all other phases are defined relative to $\varphi_{M}^{0}$.
Second, although the equations are analytical, finding a numerical solution is not straightforward, especially when considering the uncertainties associated with the measured observables. 
We therefore use the ratio $R_I^{E,M}$ in Eq.~\eqref{eq:dataRIExi} to determine the allowed boundaries for the relative magnitudes and phases in the isospin basis. 
The extracted values of $R_I^{E,M}$ and their uncertainties, as listed in Table~\ref{tab:charmonium}, determine the allowed regions for the relative magnitudes $\delta_{E,M}$ and phases $\phi_{E,M}$, which are illustrated in Figs.~\ref{fig:RIEMofJPsi} and \ref{fig:RIEMofPsi2S}.
Consistent with the expectation that isospin violation is small, we restrict the plot to $\delta_{E,M} \leq 1$.
The blue band representing $R_I^{E}$ lies above the orange band for $R_I^{M}$, reflecting the observed hierarchy $R_I^{E} > R_I^{M}$ \footnote{Previous BESIII data with lower precision suggested the opposite hierarchy, $R_I^{E} < R_I^{M}$, in both channels (see Table \ref{tab:charmonium} and Ref.~\cite{Dai:2023vsw}), underscoring the crucial importance of high-precision data for such investigations.}. 
For the $J/\psi \rightarrow \Xi\bar{\Xi}$ decay, the data provide unambiguous significance for this relation. 
However, for $\psi(2S) \rightarrow \Xi\bar{\Xi}$, the two bands exhibit some overlap, providing only marginal evidence. 
Future updated measurements of the branching ratios would be essential to clarify this situation.
Within the constraints of the allowed parameter space, a $\chi^2$ minimization is subsequently employed to obtain numerical solutions to Eqs.~\eqref{eq:GMsum}--\eqref{eq:EM4} within the allowed parameter space. Bootstrap resampling with 2000 iterations is used to propagate the uncertainties of the observables.
The resulting numerical solutions, including their uncertainties, are listed in Table~\ref{tab:G01} and presented as colored points with error bars in Figs.~\ref{fig:RIEMofJPsi} and \ref{fig:RIEMofPsi2S}.
As our primary conclusion, evidence for $\delta_{E} > \delta_{M}$ is observed in the $J/\psi$ channel at the $2\sigma$ significance level, with similar indications found in the $\psi(2S)$ channel. 
The isospin-violating magnitudes for the electric and magnetic couplings are approximately 10\% and 3\%, respectively, for both charmonium states. 
The relative phase $\phi_{M}$ is found to be consistent with zero, while the relative phase $\phi_{E}$ shows a significant deviation from zero in both decay channels. Interestingly, the value of $\varphi_{E}^{1}-\varphi_{M}^{0}$ for $\psi(2S)$ remains consistent with zero, whereas that for $J/\psi$ deviates substantially from zero, suggesting distinct decay mechanisms for the two charmonium states.

\begin{table}
    \centering
    \caption{The isoscalar $G_{E,M}^0$ and isovector $G_{E,M}^1$ couplings and their phases (in rad) for $J/\psi$ and $\psi(2S)$ decaying into $\Xi\bar{\Xi}$. }
    \begin{tabular}{ccc}
\hline\noalign{\smallskip}
channel  & $J/\psi \to \Xi \overline \Xi$  & $\psi(2S) \to \Xi \overline \Xi$   \\
\hline\noalign{\smallskip}
\hline
    $|G_M^0|$ & $(2.77\pm0.066)\times10^{-3}$ & $(2.06\pm0.043)\times10^{-3}$ \\ 
    $|G_M^1|$ & $(7.91\pm6.01)\times10^{-5}$ & $(6.21\pm3.05)\times10^{-5}$ \\
    $|G_E^0|$ & $(1.39\pm0.038)\times10^{-3}$ & $(8.09\pm1.01)\times10^{-4}$ \\
    $|G_E^1|$ & $(1.17\pm0.41)\times10^{-4}$ & $(8.66\pm3.02)\times10^{-5}$ \\
    $\varphi_{M}^{1}-\varphi_{M}^{0}$ & -0.0053$\pm$0.29 & -0.0015$\pm$0.0008 \\
    $\varphi_{E}^{0}-\varphi_{M}^{0}$ & -1.19$\pm$0.027 & -0.36$\pm$0.10 \\
    $\varphi_{E}^{1}-\varphi_{M}^{0}$ & -0.95$\pm$0.55 & 0.0025$\pm$0.0012 \\
    \hline
    $\delta_{E}$ & $0.084^{+0.030}_{-0.013}$ & 0.11 $\pm$ 0.040  \\
    $\delta_{M}$ & $0.029^{+0.022}_{-0.013}$ & $0.030^{+0.006}_{-0.015}$ \\
     \noalign{\smallskip}\hline\noalign{\smallskip}
\hline
    \end{tabular}
    \label{tab:G01}
\end{table}

\section{$S-$ and $D-$ partial waves} \label{sec:sdwave}

The studied amplitudes can be alternatively described using various formalisms, such as helicity amplitudes \cite{Buttimore:2006mq,Buttimore:2007cv}, covariant tensor formalisms \cite{Wu:2021yfv}, or by considering Dirac and Pauli couplings \cite{Barnes:2007ub}. 
In this section, we focus on an alternative decomposition in terms of $S-$ and $D-$ partial waves, which is equivalent to the electromagnetic form factor representation~\cite{Wu:2021yfv,Haidenbauer:2020wyp,Yang:2022qoy,Yang:2024iuc,Dai:2024lau}:
\bea \label{eq:GMNpw}
G_M &=& g_s + \frac{1}{\sqrt{2}} g_d \\ \label{eq:GENpw}
\frac{G_E}{\sqrt{\tau}} &=& g_s - \sqrt{2} g_d 
\eea
Note that we adopt a normalization convention different from that in the literature, chosen to express the partial branching fractions in a more transparent manner:
\bea \label{eq:widthsd}
\mcb_{\psi \to Y \bar{Y}} &=& \frac{M_{\psi}\beta}{8 \pi \Gamma_{\psi}}\lf( |g_s|^2 + |g_d|^2 \rg) \,,
\eea
In the isospin partial-wave basis, the couplings are typically written as,
\bea
g_{s,d}^{1,0} &=&  \frac{g_{s,d}^{+} \pm g_{s,d}^{-}}{2}
\eea
where the upper sign 1 (or 0) applies to the isovector (or isoscalar) case.
$g_{s,d}^{+}$ (or $g_{s,d}^{-}$) are $S-$wave and $D-$wave coupling constants of $\Xi^{0}$ (or $\Xi^{-}$), respectively.
Consequently, the left-hand sides of Eqs.~\eqref{eq:GMsum}--\eqref{eq:EM4} can be rewritten using $g_{s,d}^{\pm}$.
The complex amplitudes are further parameterized as $g^{I}_{s,d} = |g^{I}_{s,d}| \, e^{i \varphi_{s,d}^I}$.

\begin{figure}
    \centering
    \includegraphics[width=0.75\linewidth]{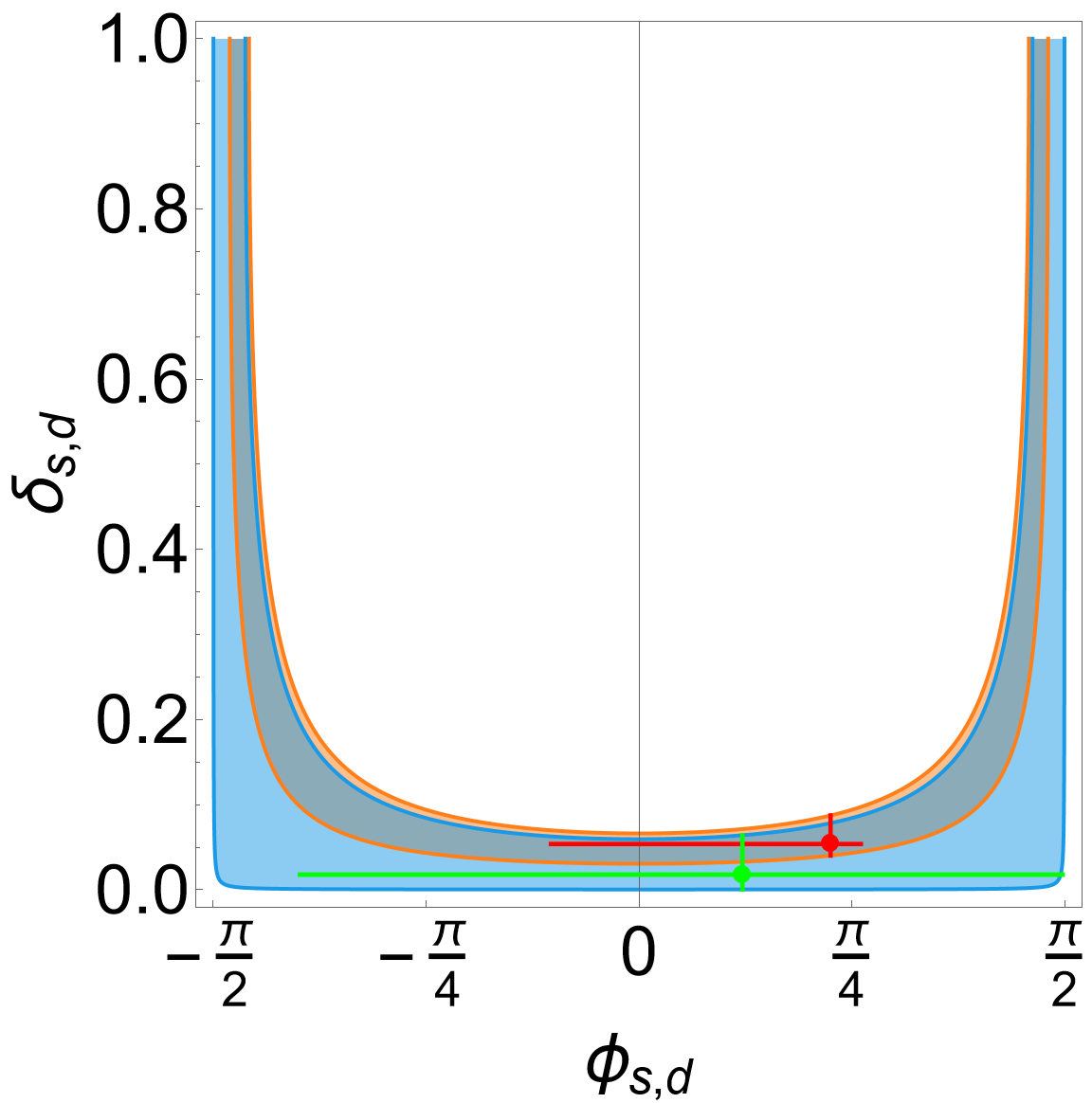}
    \caption{Allowed regions for $\delta_{s,d}$ and $\phi_{s,d}$ in $J/\psi\rightarrow\Xi\bar{\Xi}$ decays. The blue and orange bands represent constraints derived from the $R_I^S$ and $R_I^D$, respectively. The points with error bars denote the numerical solutions obtained from the $\chi^2$ minimization.}
    \label{fig:RsdofJPsi}
\end{figure}

\begin{figure}
    \centering
    \includegraphics[width=0.75\linewidth]{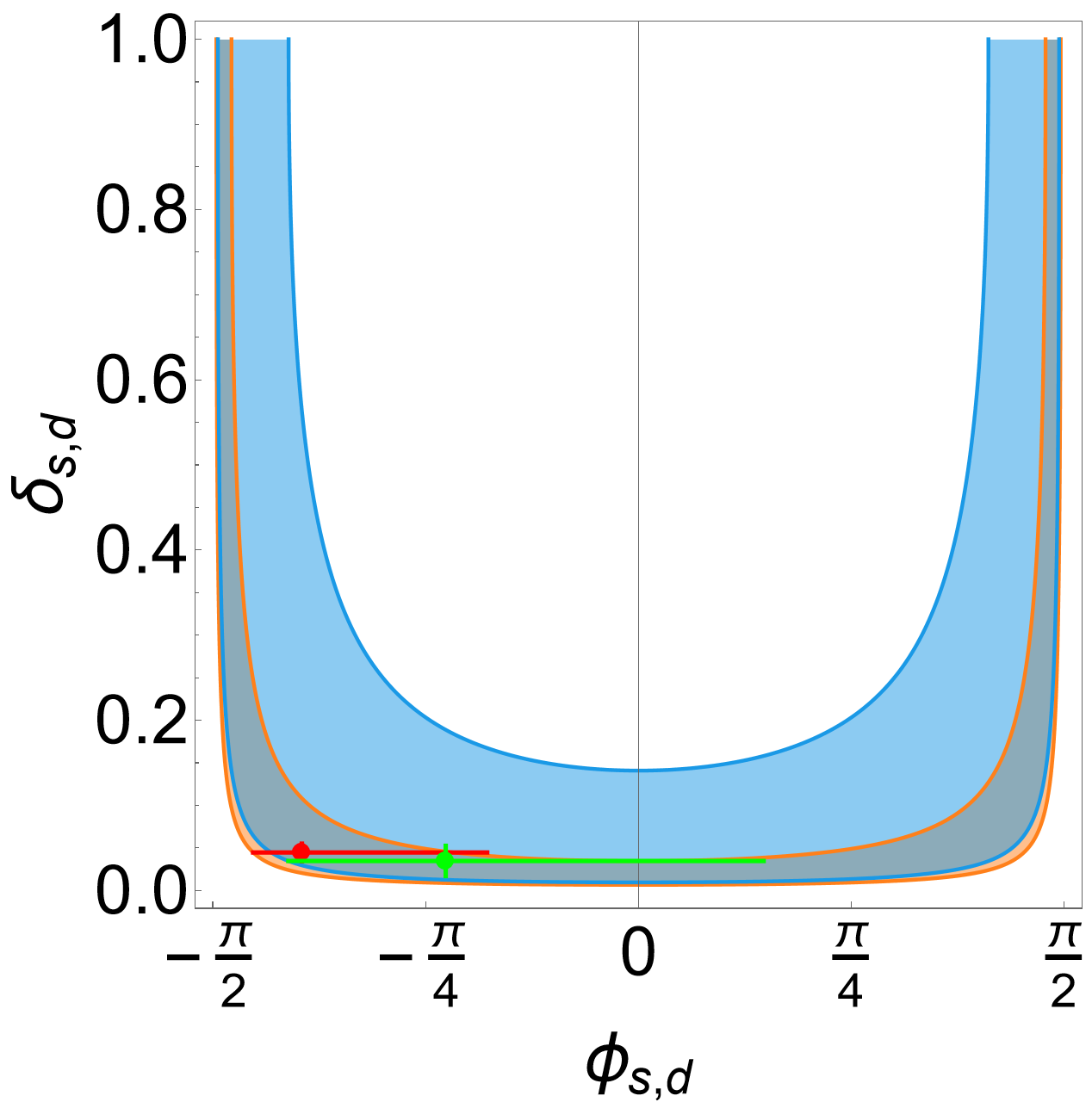}
    \caption{Same as Fig. \ref{fig:RsdofJPsi} but for $\psi(2S) \rightarrow \Xi\bar{\Xi}$ decay..}
    \label{fig:RsdofPsi2s}
\end{figure}

Following the same procedure as in Sec.~\ref{sec:isodecomp}, a partial-wave relevant ratio is introduced to bound the extent of isospin violation:
\be \label{eq:dataRIsd}
R_I^{S,D} = \frac{|g_{s,d}^{+}|^2 - |g_{s,d}^{-}|^2}{|g_{s,d}^{+}|^2 + |g_{s,d}^{-}|^2} = \frac{2\delta_{s,d}\cos{\phi_{s,d}}}{1+\delta_{s,d}^2}
\ee
where $\delta_{s,d} = |g_{s,d}^1/g_{s,d}^0|$ is relative magnitude and $\phi_{s,d} = \varphi_{s,d}^1-\varphi_{s,d}^0$ is relative phase between isovector and isoscalar couplings.
The allowed regions for these parameters are shown in Figs.~\ref{fig:RsdofJPsi} and \ref{fig:RsdofPsi2s}, where the final solutions from the $\chi^2$ minimization are indicated by colored points with error bars.
The corresponding numerical results are summarized in Table~\ref{tab:sdwave}.
To account for the arbitrary global phase, $\varphi_{s}^{0}$ is fixed at zero during the numerical search.

At current precision, the data suggest that isospin violation in the $S$-wave is of a similar magnitude to that in the $D$-wave.
The large errors in the extracted phases preclude any definitive conclusions at this stage, highlighting the need for more precise data in future experiments.

\begin{table}
    \centering
    \caption{The coupling constant and phases (in rad) of $S-$wave and $D-$wave for $J/\psi$ and $\psi(2S)$ decaying into $\Xi\bar{\Xi}$.}
    \begin{tabular}{ccc}
\hline\noalign{\smallskip}
channel  & $J/\psi \to \Xi \overline \Xi$  & $\psi(2S) \to \Xi \overline \Xi$   \\
\hline\noalign{\smallskip}
\hline
 $|g_s^0|$ & $(1.97\pm0.15)\times10^{-3}$ & $(15.04\pm0.28)\times10^{-4}$ \\ 
 $|g_s^1|$ & $(1.06\pm3.99))\times10^{-4}$ & $(6.72\pm1.27)\times10^{-5}$ \\
 $|g_d^0|$ & $(12.07\pm0.64)\times10^{-4}$ & $(7.99\pm0.28)\times10^{-4}$ \\
 $|g_d^1|$ & $(2.14\pm18.21)\times10^{-5}$ & $(2.77\pm1.14)\times10^{-5}$ \\
 $\varphi_{s}^{1}-\varphi_{s}^{0}$ & 0.71$\pm$1.03 & -1.24$\pm$0.34 \\
 $\varphi_{d}^{0}-\varphi_{s}^{0}$ & -1.78$\pm$1.19 & -1.18$\pm$0.05 \\
 $\varphi_{d}^{1}-\varphi_{s}^{0}$ & 0.98$\pm$1.12 & -1.90$\pm$1.21\\
    \hline
    $\delta_{s}$ & $0.054_{-0.014}^{+0.034}$  & 0.045 $\pm$ 0.010  \\
    $\delta_{d}$ & $0.018_{-0.017}^{+0.046}$  & 0.035 $\pm$ 0.018  \\
\noalign{\smallskip}\hline\noalign{\smallskip}
\hline
    \end{tabular}
    \label{tab:sdwave}
\end{table}

\section{Summary and Perspective}\label{sec:sum}

As established in earlier work~\cite{Dai:2023vsw}, the relative magnitudes of the isoscalar and isovector components in baryon-antibaryon pairs produced via electron-positron annihilation can be unambiguously determined using a complete set of observables from the relevant isospin channels. 
Although this framework has been applied to nucleon and hyperon electromagnetic form factors through differential cross-section analysis, and to charmonium decays by incorporating isospin violation, previous analyses were limited by large uncertainties and the absence of polarization observables. 
With the accumulation of high-statistics data and reduced uncertainties, it is now possible to perform a more accurate, model-independent analysis based on the isospin decomposition of vector charmonium decays into hyperon-antihyperon pairs. 
This paper is devoted to outlining the full theoretical framework for isospin decomposition in terms of electromagnetic couplings and applying it to the available experimental data for $J/\psi$ and $\psi(2S)$ decays into $\Xi\bar{\Xi}$ pairs.

Our analysis demonstrates that the isospin violations are predominantly driven by the electric couplings (approximately 10\%), whereas the magnetic contributions are significantly smaller (about 3\%). This finding is particularly interesting, as the QCD factorization framework previously predicted that isospin violations would be more sensitive to the baryon magnetic form factors~\cite{Claudson:1981fj}.
The isospin violations in the $S$- and $D$-wave components are of comparable magnitude, each at the level of several percent.
Our results contribute to the understanding of the decay mechanisms of charmonia into baryon-antibaryon pairs, while potentially offering insights into the quantum entanglement inherent in these systems \cite{Li:2026bkf,Zhang:2026wvn,Zhang:2026nwm}.

Our framework is well-suited for the isospin decomposition of nucleon-antinucleon system~\cite{BESIII:2012ion,BESIII:2018flj,BESIII:2026nmr}. For vector charmonia decay, preliminary analysis indicates that isospin violation in the electric coupling is more pronounced~\cite{Dai:2023vsw}. 
By utilizing recently developed techniques~\cite{BESIII:2026yyv,BESIII:2026lkl} to measure proton polarization, one can effectively validate the current conclusions and provide more stringent constraints on the isospin-violating amplitudes.
In contrast, measuring $\Delta \Phi$ for nucleon electromagnetic form factors at any energies with high precision is more challenging due to the high statistics required by polarization measurements. 
Alternatively, dispersion-theoretic approaches that globally fit both spacelike and timelike data provide valuable information on these phases~\cite{Lin:2021umz,Lin:2021xrc}.

Despite the increased complexity of the decomposition for isospin triplets, our framework is applicable to the $J/\psi, \psi(2S) \rightarrow \Sigma^{\pm} \bar{\Sigma}^{\mp}$ final states~\cite{BES:2008hwe,BESIII:2021wkr,BESIII:2024dmr,BESIII:2025jxt,BESIII:2020fqg,BESIII:2023sgt}. Nevertheless, a complete isospin analysis is currently hindered by the lack of experimental data for the $\Sigma^-\bar{\Sigma}^+$ channel~\cite{BESIII:2022upt}.
The potentially more pronounced isospin-violating effects in these channels, estimated at approximately 30\%~\cite{Dai:2023vsw}, merit further attention.

\bigskip
\begin{acknowledgments}

We are grateful to Xiongfei Wang for useful discussions. This work is supported by the National Natural Science Foundation of China  (Grant No. 12547111 and 12165022), the National Key R\&D Program of China under Grant No. 2023YFA1606703, and Yunnan Fundamental
Research Project under Contract No. 202301AT070162.).

\end{acknowledgments}

\bigskip

\bibliography{NEFF_all.bib}

\end{document}